\documentclass[reprint, amsmath, amssymb, aps, prl]{revtex4-2}
\usepackage{graphicx, dcolumn, bm, bbm,amsmath, amssymb, xcolor, enumitem, float}
\usepackage[spaces]{grffile}
\usepackage{hyperref}
\allowdisplaybreaks

\let\tempvec\vec
\renewcommand{\vec}[1]{\tempvec{{}#1}}
\let\tempbar\bar
\renewcommand{\bar}[1]{\tempbar{{}#1}}

\begin{document}
	\title{Geometrically frustrated, mechanical metamaterial membranes: \\
Large-scale stress accumulation and size-selective assembly}


	\author{Michael Wang$^{1,}$}
		\email{mwang@mail.pse.umass.edu}
    
	\author{Sourav Roy$^{2,}$}
        \email{sroy08@syr.edu}
      
	\author{Christian Santangelo$^{2,}$}
        \email{cdsantan@syr.edu}
	
    \author{Gregory Grason$^{1,}$}
        \email{grason@umass.edu}
      
	\affiliation{$^1$Department of Polymer Science and Engineering, University of Massachusetts, Amherst, MA 01003 \\
 $^2$Department of Physics, Syracuse University, NY 13210}

	\begin{abstract}
        We study the effect of geometric frustration on dilational mechanical metamaterial membranes.  While shape frustrated elastic plates can only accommodate non-zero Gaussian curvature up to size scales that ultimately vanish with their elastic thickness, we show that frustrated {\it metamembranes} accumulate hyperbolic curvatures up to mesoscopic length scales that are ultimately independent of the size of their microscopic constituents.  A continuum elastic theory and discrete numerical model describe the size-dependent shape and internal stresses of axisymmetric, trumpet-like frustrated metamembranes, revealing a non-trivial crossover to a much weaker power-law growth in elastic strain energy with size than in frustrated elastic membranes.  We study a consequence of this for the self-limiting assembly thermodynamics of frustrated trumpets, showing a several-fold increase the size range of self-limitation of metamembranes relative to elastic membranes.  
	\end{abstract}
 
	\maketitle


        Geometric frustration occurs when a locally preferred structural motif cannot be realized globally \cite{Sadoc:2006}.  
        In soft matter systems, from liquid crystals \cite{Wright:1989,Niv:2018,Selinger:2022} to colloids \cite{Meng:2014,Irvine:2010,Li:2019} to filamentous assemblies \cite{Bruss:2012,Panaitescu:2018,Atkinson:2021}, geometric frustration gives rise to the accumulation of elastic distortions at sizes much larger than the microscopic material elements \cite{Grason:2016, Meiri:2021, Meiri:2022, Hackney:2023}.  Geometrically-frustrated elastic sheets are a prototypical and particularly well-studied class of such systems \cite{Sharon:2010}, with broad applications to engineered and natural materials ranging from nano- to macro-scales.  Broadly defined, frustration in elastic membranes derives from the incompatibility between the preferred metric (i.e.\ intrinsic geometry) and the preferred  curvatures (i.e.\ extrinsic geometry) of the membrane \cite{Gemmer:2013}.  For example, the differential growth \cite{Liang:2007} or swelling \cite{Klein:2007, Kim:2012} of a sheet that is uniform through its thickness gives rise to a preferred non-zero Gaussian curvature, which competes with the energy cost of bending the sheet to the preferred curvature. These so-called {\it non-Euclidean plates}  exhibit a complex spectrum of compromised equilibria, such as the undulating shapes of leaves \cite{Liang:2007} and torn plastic wrap \cite{Sharon:2007}. For so-called {\it hyperbolic Euclidean shells}, on the other hand, the bending energy prefers a nonzero Gaussian curvature but competes with a stretching energy away from intrinsically flat metric relationships \cite{Armon:2011}.
        \begin{figure}
            \centering
            \includegraphics[width=\columnwidth]{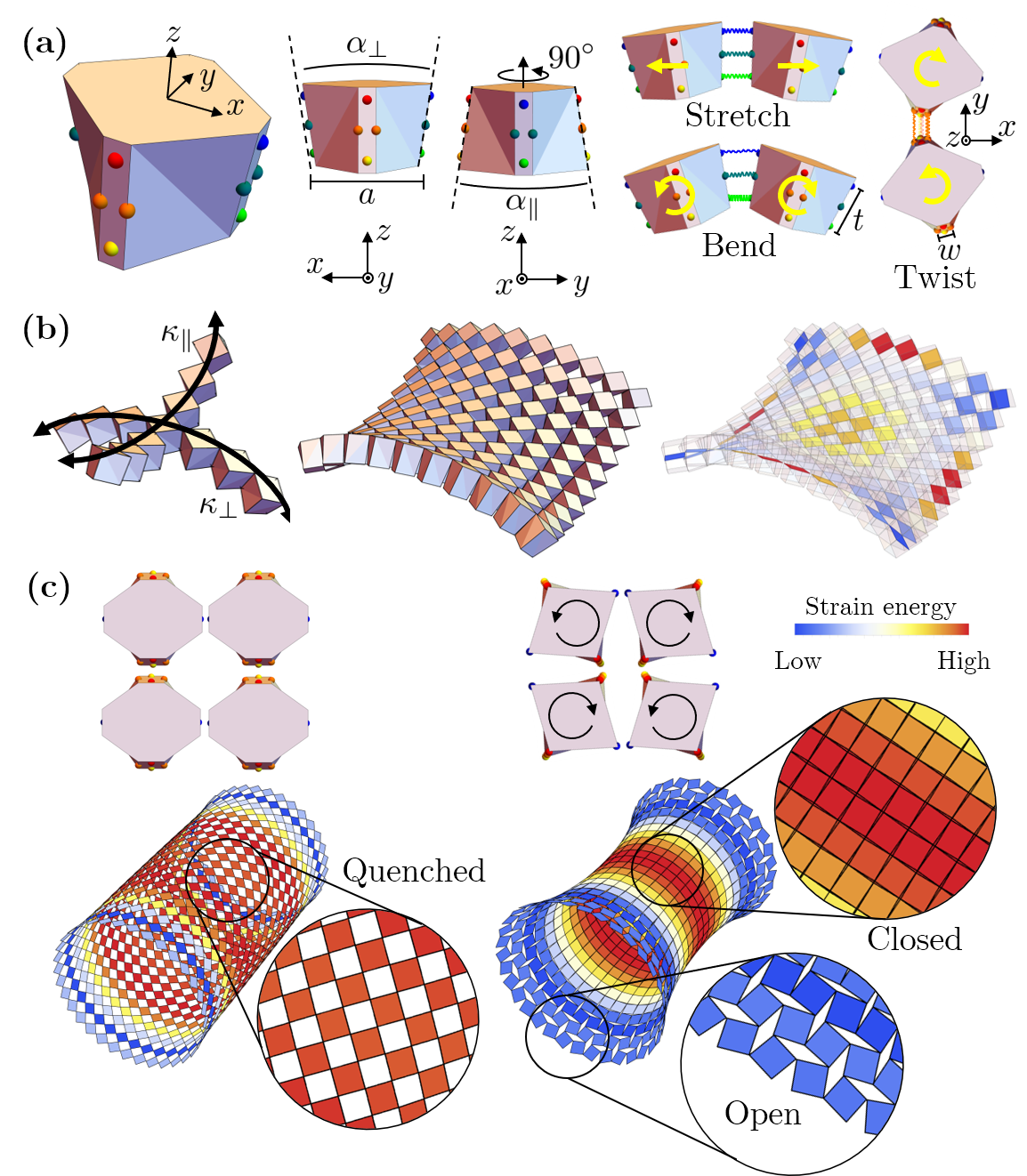}
            \caption{``Cuboidal" subunit model of hyperbolically frustrated membranes.  \textbf{(a)} Subunit shape and attractive sites (blue, green, and red spheres) and modes of bond deformations.  \textbf{(b)} Preferred curvature of subunit rows (left) and 2D sheet assembly, favoring negative Gaussian curvature (center), and giving rise to elastic energy gradients, highlighted in (square) cross-sectional cuts (right).  \textbf{(c)} ``Trumpet'' metamembranes  with ({\it floppy}) and without (``quenched'' or {\it stiff}) a soft dilational mode.}
            \label{fig:intro}
        \end{figure}

        The competition between metric (stretching) or shape (bending) in frustrated elastic membranes depends sensitively on lateral dimensions, giving rise to remarkably non-linear, shape-adaptive behavior.  In hyperbolic Euclidean shells, narrow sheets accommodate the stretching cost needed to adopt the preferred Gaussian curvature, while wider sheets flatten to expel it to a narrow boundary layer at their edges \cite{Armon:2014, Grossman:2016}.  This size-dependent frustration drives the shape transition between helicoids to spiral ribbons observed in diverse systems such as bauhinia seed pods \cite{Armon:2011}, inorganic nanosheet ribbons \cite{Serafin:2021, Monego:2024} and crystalline membranes of chiral amphiphiles \cite{Ghafouri:2005,Armon:2014,Zhang:2019,Selinger:2004}. The size scale at which the transition occurs, what we call the {\it flattening size}, $\ell_{\rm flat}$~\cite{Hagan:2021}, is set by the ratio of the elastic cost of bending to stretching in elastic sheets, and is known to vanish as the microscopic thickness, $t$ of the sheet goes to zero ~\cite{Armon:2014, Efrati:2009}.  This has important implications for {\it self-assembled frustrated membranes}, since the nature of internal stress accumulation with increasing size determines the range over which assembly thermodynamics can ``sense'' and limit the lateral width \cite{Ghafouri:2005,Armon:2014,Hagan:2021}. The size range $\ell_{\rm flat}$ and ability of frustrated elastic membranes to accumulate Gaussian curvature is especially limited \cite{Hall:2023}.  
        
        In this Letter, we study the effect of geometric frustration in a mechanical metamaterial~\cite{bertoldi:2017, santangelo:2017, Zheng:2023} membrane, dubbed a {\it metamembrane}, characterized by a bulk dilation deformation~\cite{sun:2012} that is floppy (i.e.\ zero-energy) in the absence of frustration. The existence of the floppy mode associated with the internal buckling of the metamembrane, completely reshapes both the nature and size-sensitivity of stress accumulation. We use a combination of continuum theory and numerical simulations of 2D ordered metamembranes formed from subunits whose shape favors locally negative Gaussian curvature, with a corner-binding geometry that permits an auxetic dilational mode (Fig. \ref{fig:intro}).  In the absence of geometric frustration, such structures are known as conformal metamaterials ~\cite{Zheng:2022, Czajkowski:2022}, and their 2D (in-plane) mechanics has been the focus of studies of their exotic topological mechanics ~\cite{paulose:2015, Mao:2018} as well as engineered metamaterial responses ~\cite{Konakovic:2016, Roy:2023}.  Here, we show that the introduction of geometric frustration through a preference for hyperbolic shapes internally stresses the otherwise floppy metamembrane.  The local dilational modes screen the elastic costs of frustration, ~\cite{Livne:2023}effectively allowing the metamembrane to accumulate a much larger amount of Gaussian curvature over a much larger range of sizes.  Unlike elastic membranes, this effect does not vanish in the continuum limit of vanishing subunit size.  We demonstrate that this leads to a several fold increase in the thermodynamically self-limiting sizes of frustrated tubules that can be self-assembled from discretely, ill-fitting subunits.  This opens up access to fundamentally distinct dimensions of structural control in programmable self-assemblies.

        \begin{figure*}
            \centering
            \includegraphics[width=\textwidth]{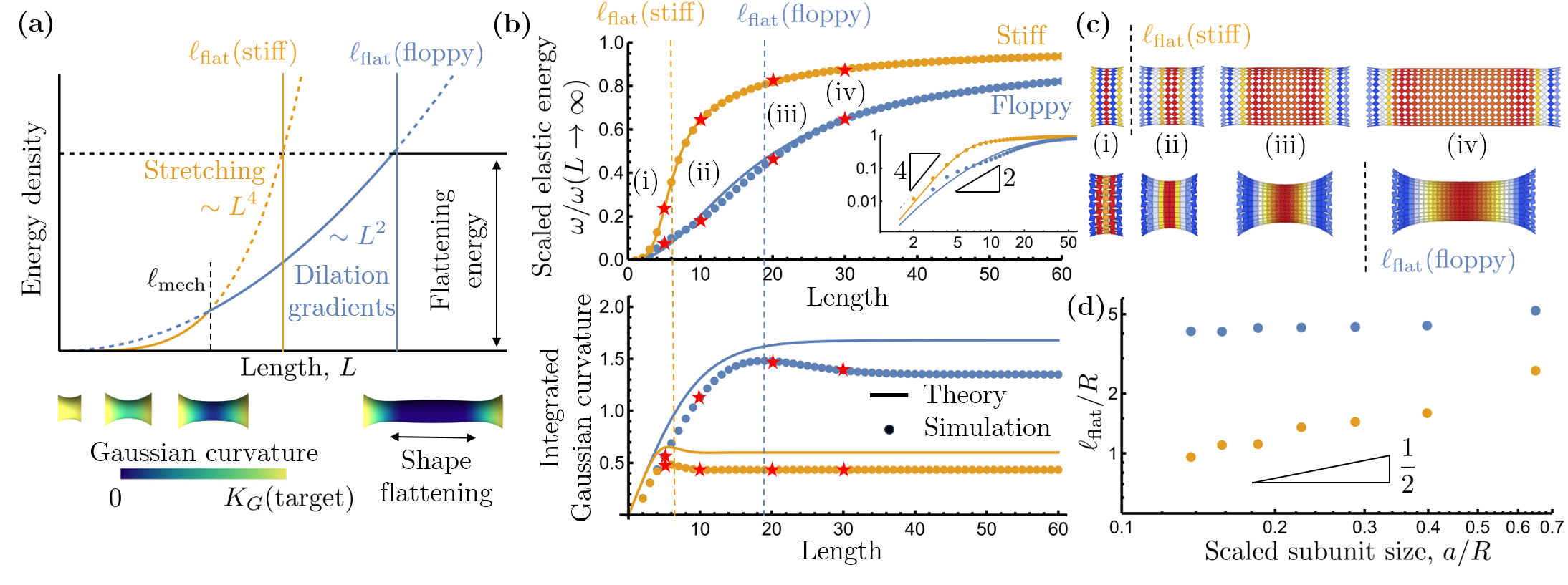}
            \caption{Energetics and shape of frustrated trumpets.  \textbf{(a)} Schematic for the energetic regimes (stretching, dilations, and flattening) of frustrated trumpets.  \textbf{(b)} Elastic energy per unit area $\omega(L)={\cal W}(L)/A$ and integrated Gaussian curvature as functions of trumpet length for the geometry $\alpha_{\perp}=\pi/28$ and $\alpha_{\parallel}=(3/4)\alpha_{\perp}$, normalized by the flattening limit ($L \to \infty$).  Solid lines are mappings to the continuum theory for a stiff (orange) mechanism and a floppy (blue) mechanism with $\ell_{\rm mech}/\ell_{\rm flat}(\rm floppy)\approx0.25$ and $C_{\rm mech}=0$.  Vertical dashed lines indicated the rough onset of flattening for trumpets with stiff (orange) and floppy (blue) dilational modes.  \textbf{(c)} Ground state structures from \textbf{(b)}, showing flattening at roughly 6 and 19 for stiff and floppy trumpets, respectively.  \textbf{(d)} Flattening length of trumpets versus subunit size (for $\kappa_{\parallel}/\kappa_{\perp}=3/4$).}
            \label{fig:energy}
        \end{figure*}

    We consider a discrete-subunit model of hyperbolically frustrated metamembranes,(Fig.~\ref{fig:intro}), composed of 2D arrays of corner-binding, rigid, quasi-cuboidal units.  In the absence of frustration, these structures exhibit a square-twist mechanism, in which counter-rotation of alternating cubes leads to 2D auxetic, dilational mechanical metamaterial behavior.  To introduce frustration, opposing corner edges are flared by angles $+\alpha_{\parallel}$ and $-\alpha_{\perp}$ (Fig.~\ref{fig:intro}a).  Ideal edge-to-edge binding along orthogonal rows gives rise to respective curvatures $\kappa_\parallel \simeq \alpha_{\parallel}/a$ and $\kappa_\perp \simeq \alpha_{\perp}/a$, where $a$ is the diagonal subunit width.  Edges between bound units are held together by four attractive sites---two vertically separated by $t$ and two horizontally separated by $w$---with hookean springs of stiffness $k$ between the sites.  Hence, as illustrated in Fig~\ref{fig:intro}a, distortions of these bonds determine the stiffness of three key modes of inter-unit mechanics: center-to-center stretching ($\propto k$); inter-unit bending ($\propto kt^2$); and inter-unit twisting ($\propto kw^2$).  
    
    Pairwise mechanics determines the collective mechanics of 2D sheets of these cuboidal units (Fig.~\ref{fig:intro}b), for an example targeting minimal surface shapes ($\kappa_\parallel=-\kappa_\perp$) showing gradients of strain energy in its elastic ground state.  Notably, the inter-particle twist stiffness dramatically impacts the ability of frustrated metamembranes to accommodate non-zero Gaussian curvature.  This is shown for the tubular assemblies in Fig.~\ref{fig:intro}c, where the hoop direction prefers a self-closure radius $R=\kappa_\perp^{-1}$, while the axial curvature prefers a {\it trumpet-like} flare~\cite{Tyukodi:2022}.  When the twist mode is {\it stiff} (e.g.\ $w \approx t$), the mechanism is effectively quenched and the equilibrium shape is only weakly perturbed from a uniform cylindrical shape near the edges.  In contrast, when the twist mode is {\it floppy} (e.g.\ $w =0$), equilibrium shapes exhibit a hyperbolic curvature over the bulk of the assembly, as well as larger scale gradients in strain energy.

    We can understand the nature of shape and stress accumulation in frustrated metamembranes by way of a continuum theory, which combines elements of a F\"oppl-van K\'arman like description of dilational metamaterial sheets~\cite{Roy:2023} with those of the theory of hyperbolic Euclidean shells~\cite{Armon:2014} (see SI for derivation).  This model is described by an elastic energy ${\cal W}$ that is a function of elastic strains $\bm{\varepsilon}$ within the 2D array, the curvature tensor $\bm{\kappa}$ of its out-of-plane shape, and a scalar field $\Omega$ characterizing the local actuation of the square-twist dilational mechanism.  The energy takes the form
    \begin{multline}
        {\cal W}({\bm \varepsilon},{\bm \kappa}, \Omega ) = \int dA\left\{\frac{1}{2}\Big[ (\lambda +  \mu) \big( \varepsilon_{\rm dil} - \Omega\big)^2 + 2 \mu {\bm \varepsilon}^2_{\rm dev} \Big]\right. \\ \left.+ \frac{B}{2} \big({\bm \kappa} - \bar{{\bm \kappa}}\big)^2 + \frac{C_{\rm mech}}{2} |\Omega |^2+ \frac{C_\nabla}{2}|\nabla \Omega |^2 \right\},
    \end{multline}
    where $dA$ is the surface element.  The bulk modulus $(\lambda + \mu)$ couples the dilational strain $\varepsilon_{\rm dil} = {\rm Tr} {\bm \varepsilon}$, to the twist mechanism, while deviatoric strains ${\bm \varepsilon}_{\rm dev} ={\bm \varepsilon} - \mathbbm{1}\varepsilon_{\rm dil}$, are uncoupled from the mechanisms and penalized by a shear modulus $\mu$.  The third term describes the bending cost with modulus $B$ of a membrane penalizing deviation from its {\it target curvatures}
    \begin{equation}
        \bar{{\bm \kappa}} = \left( \begin{array}{cc}  \kappa_\parallel & 0 \\ 0 &- \kappa_\perp  \end{array} \right).
    \end{equation}
    The last two terms describe the costs for uniform and non-uniform actuation of the twist mechanism, where $\nabla \Omega$ is its in-plane gradients.  It is straightforward to show (see SI) that the moduli can be related to the subunit model (Fig.~\ref{fig:intro}), with $\lambda \propto \mu \propto k$, $B\propto k t^2$, $C_{\rm mech}\propto k (w/a)^2$ and $C_\nabla \propto ka^2$, which are consistent ~\cite{Czajkowski:2022, Zheng:2023} for planar dilational metamaterials.  

    While it is straightforward to solve this continuum theory for the shape equilibria of axisymmetric trumpets (see SI), simple energy arguments suffice to explain the differences in how elastic energy density, $\omega \equiv {\cal W}/A$, accumulates with length $L$ for {\it stiff} ($C_{\rm mech} \approx k$) versus {\it floppy} ($C_{\rm mech}=0$) metamembrane trumpets.  For the stiff case ($C_{\rm mech}\approx k$), the mechanism is quenched ($\Omega=0$)  at all relevant scales, and the bending energy dominates to maintain a target hyperbolic shape with Gaussian curvature $K_{\rm G} \approx- \kappa_\perp \kappa_\parallel$.  Gauss' theorem prescribes the dilational strain gradients needed accommodate the non-Euclidean geometry $\nabla^2 \varepsilon_{\rm dil} \simeq - K_{\rm G}$, which is solved by the quadratic growth of strain with axial distance $z$ from the mid-point of the trumpet, $\varepsilon_{\rm dil}(z) = \varepsilon_{\rm dil}(0)-K_{\rm G}z^2/2$, and a stretching energy density that grows as $\omega(L \to 0)\sim Y K_{\rm G}^2 L^4$, where $Y\propto 4 \mu (\lambda+\mu)/(\lambda+2\mu)$ is the (bare) 2D Young's modulus.  This quartic growth with length (Fig.~\ref{fig:energy}a) proceeds until trumpets reach a size where stretching exceeds the finite energy density $\omega(L \to \infty)\sim \frac{B}{2} \kappa_\parallel^2$ to unbend, or {\it flatten}, the shape along the axial direction.  This size is called the {\it flattening length} $\ell_{\rm flat}({\rm stiff}) = (B/Y)^{1/4} \kappa_\perp^{-1/2} \propto \sqrt{t R} $, which is the well known narrow size of the curved boundary layer in frustrated membranes~\cite{Efrati:2009}.   
    
    For the floppy case ($C_{\rm mech}=0$), the twist mechanism is free to screen dilational strains via $\Omega(z) \simeq \varepsilon_{\rm dil}(z)$, leading to a residual cost for twist gradients $\partial_z \Omega\approx -K_{\rm G} z$ that grows only quadratically with size as $\omega\sim C_{\nabla} K_{\rm G}^2 L^2$.  Dlational screening by twist (Fig.~\ref{fig:energy}a) occurs for $L \gtrsim \ell_{\rm mech} = \sqrt{C_{\nabla}/Y}\propto a$, beyond which the soft accumulation of elastic energy leads to a larger flattening size  $\ell_{\rm flat}({\rm floppy}) = (B/C_{\nabla})^{1/2} \kappa_\perp^{-1} \propto (t/a) R $.  It is interesting to observe that the screening of dilational strains in metamembranes leads to a quadratic power-law accumulation of frustration cost, akin to orientational strains in frustrated {\it liquid crystalline} membranes~\cite{Nelson:1987, Meiri:2022}.

    \begin{figure*}
        \centering
        \includegraphics[width=\textwidth]{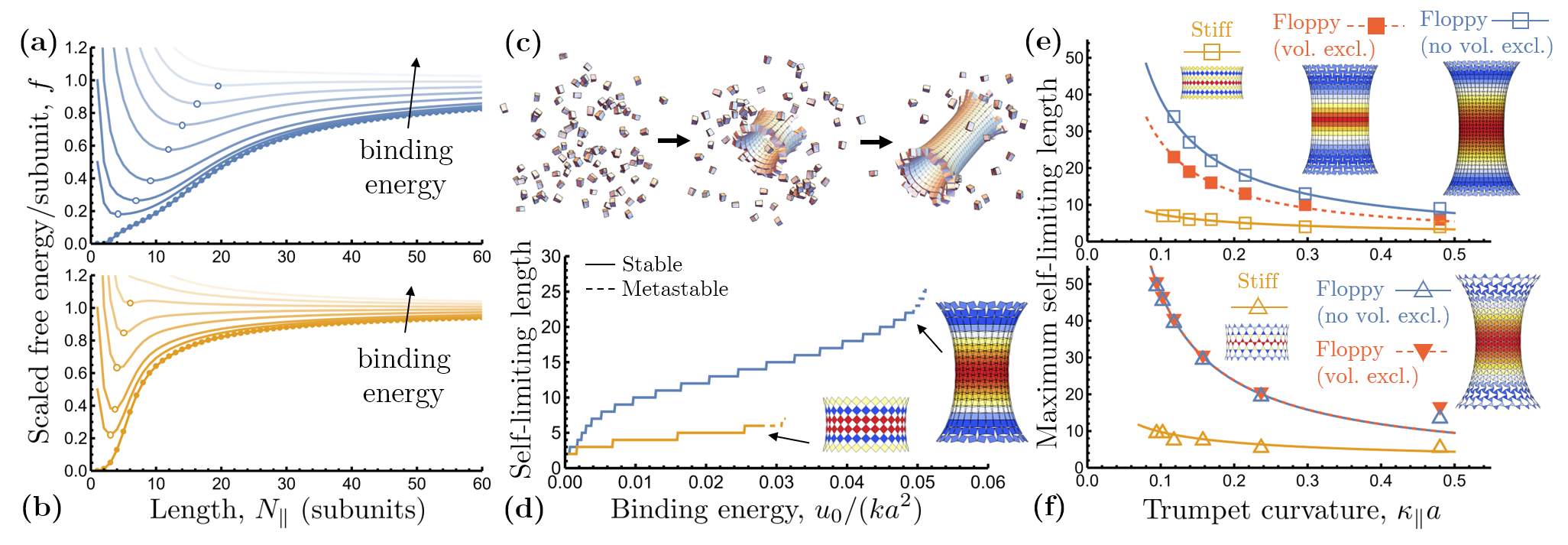}
        \caption{Free energy per subunit, normalized by the unlimited flattened ($L \to \infty$) case, for the assembly of \textbf{(a)} {\it stiff} and \textbf{(b)} {\it floppy} axisymmetric trumpets, shown schematically in \textbf{(c)}. Target curvatures $\kappa_{\perp}=\pi/28$ and $\kappa_{\parallel}=(3/4)\kappa_{\perp}$, with increasing transparency indicating increasing binding free energy, $u_0$, and open circles marking the self-limiting sizes.  \textbf{(d)} Self-limiting length versus binding energy for stiff and floppy trumpets, with the largest equilibrium structure shown for each.  Dashed lines show the metastable local minima (i.e.\ thermodynamically unlimited regime). Maximum self-limiting length for trumpets with \textbf{(e)} square-twist and \textbf{(f)} Kagome metastructures, as functions of the target trumpet curvature $\kappa_{\parallel}$ (with fixed ratio $\kappa_{\parallel}/\kappa_{\perp} = 3/4$).  Example structures shown have axial curvatures $\kappa_{\parallel}a\approx0.12$ for square-twist and $\kappa_{\parallel}a\approx0.10$ for Kagome.  Solid and dashed lines show the scalings $(\kappa_{\parallel}a)^{-1/2}$ for a stiff mode (orange) and $(\kappa_{\parallel}a)^{-1}$ for a floppy mode (red and blue).}
        \label{fig:max size}
    \end{figure*}

    We test this accumulation of elastic energy density with length $N_\parallel$, the number of subunits along the axial direction, for stiff versus floppy frustrated trumpets with a fixed geometry $\kappa_\parallel=(3/4)\kappa_\perp$ in Fig.~\ref{fig:energy}b (top).  The comparison between continuum theory and discrete metamembranes shows a non-linear softening of elastic energy growth for floppy trumpets relative to stiff ones, with both smoothly saturating as $L \to \infty$.  The inset shows an obvious $L^4$ growth of stretching energy for stiff trumpets, while for floppy trumpets the discrete subunit model suggests a power-law slightly weaker than the predicted $L^2$ growth.  We characterize the range of frustration accumulation in terms of the integrated Gaussian curvature (Fig.~\ref{fig:energy}b) (bottom), which saturates just beyond a weak local maximum, marked by $\ell_{flat}$.  Not only is the size range of flattening much larger for floppy trumpets compared to stiff ones, but so too is the total accumulation of non-zero Gaussian curvature in the $L \to \infty$ limit (Fig.~\ref{fig:energy}c). We highlight the fact that, while $\ell_{\rm flat}({\rm stiff}) \sim t^{1/2}$ implies that the size scale of the $K_{\rm G} \neq 0$ zone vanishes as $a \to 0$ (for fixed aspect ratio $a/t$), we find that the (larger) value of $\ell_{\rm flat}({\rm floppy})$ is independent of microscopic subunit size in this same limit (Fig.~\ref{fig:energy}d).

   This distinction can be recast in terms of the so-called F\"oppl-van K\'arman number~\cite{Witten:2007}, which measures the relative importance of stretching versus bending in an elastic sheet. In the case of the trumpet in standard 2D elasticity, this takes the form $Y \kappa^2 L^4 /B \propto (\kappa L)^2 (L/t)^2$, which diverges in the infinitely thin sheet limit (for fixed mesoscopic dimensions $\kappa L$), indicating that stretching energy generic favors expulsion of $K_{\rm G}$ at large sizes. In the case of a metamembrane, the corresponding F\"oppl-van K\'arman number, $(\kappa L)^2 (t/ a)^2\propto(\kappa L)^2$, implies that the mesoscopic dimensions alone determine the ability to absorb Gaussian curvature.

        The much softer accumulation of elastic energy with size in frustrated floppy metamembranes, has important consequences on their ability to exhibit thermodynamic self-limiting states.  In general, equilibrium self-limitation derives from a local minimum in the per subunit assembly free energy. This behavior has recently been reported in simulations of models of frustrated crystalline membranes~\cite{Tyukodi:2022}, notably inspired by a recently developed class of shape-programmable DNA origami building blocks~\cite{Sigl:2021, Hayakawa:2022}.  However the narrow range of elastic energy accumulation limits the self-limiting sizes of these trumpet-like elastic membranes to roughly $\lesssim 4-8$ subunits in width~\cite{Hall:2023, Tyukodi:2022}.  Here, we consider the per subunit energy of defect-free, axisymmetric metamembrane trumpet assemblies, shown to describe relevant assembly states in finite-$T$ simulations \cite{Tyukodi:2022}, which is parameterized by a binding free energy $-u_0$ per bound corner,
        \begin{equation}
            f(N_\parallel) = -2 u_0 + \frac{u_0 N_\perp}{N_\parallel}+\omega(N_\parallel) ,
            \label{eq: f}
        \end{equation} 
        for trumpets with a fixed (ideal) hoop radius $N_\perp=R/a$.
        The per subunit cohesive cost $u_0 N_\perp/N_\parallel$ of the open boundary in the second term drives the assembly to larger lengths, which competes with the monotonically increasing per subunit elastic cost of frustration $\omega$.  Figs.~\ref{fig:max size}a-b show plots of the normalized per subunit free energy for stiff and floppy axisymmetric trumpets predicted by Eq.\ (\ref{eq: f}), for an increasing sequence of binding strengths.  For sufficiently small $u_0$, $f(N_\parallel)$ has minima at finite sizes $N^*_\parallel$, indicating a thermodynamically favored self-limiting length.  Increasing binding strength increases that self-limiting size as well as the value of $f(N^*_\parallel)$.  For sufficiently large $u_0$, the flattening of $\omega(N_\parallel)$ at large sizes leads either to $f(N^*_\parallel) > f(N_\parallel \to \infty)$ (i.e. metastable finite length) or the disappearance of a finite-$N_\parallel$ local minimum, both of which indicate the thermodynamically unlimited growth of flattened trumpet assemblies.  The dependence of the equilibrium self-limiting lengths on cohesive binding for stiff and floppy assemblies is shown in Fig. ~\ref{fig:max size}d.  Notably, the softer and longer range of elastic energy accumulation in floppy assemblies results in a significantly enhanced range of self-limitation.  For the same target curvatures, the maximal self-limiting lengths for floppy trumpets extend up to 22 subunits in comparison to the modest upper limit of 6 subunits for stiff trumpets (Figs.~\ref{fig:max size}e-f).     

        In the absence of any excluded volume interactions, the maximal self-limiting size, as a function of the target curvature $\kappa_\parallel$ (for fixed $\kappa_\parallel/\kappa_\perp$), of floppy trumpets is $\sim 4-5$ fold larger than stiff assemblies (Fig.~\ref{fig:max size}e).  Moreover, we observe that the maximal length grows much more rapidly as the target curvature tends to zero for floppy trumpets ($\sim \kappa_{\parallel}^{-1}$) than for stiff trumpets ($\sim \kappa_{\parallel}^{-1/2}$), which for fixed ratio $\kappa_\parallel/\kappa_\perp$, is consistent with the intuitive notion that the upper self-limiting size is proportional to flattening size, which exhibits the same scaling.  At the lower curvatures, the maximal-size structures are composed of 1360 subunits in floppy trumpets, compared to only 280 subunits in stiff trumpets. 
        
        It can be shown (see SI) that the ratio $\kappa_\parallel/\kappa_\perp$ determines the magnitude of twist $\psi(0)$ at the center of trumpets needed to screen dilational strains in hyperbolic trumpets.  For sufficiently large $\kappa_\parallel/\kappa_\perp$, dilational screening exceeds the maximal twist range ($\pi/4$) accessible for square-twist mechanisms, leading to subunit overlap at the center of floppy trumpets (Fig.~\ref{fig:max size}e).  The inclusion of excluded volume interactions (see SI) prevents overlap between cuboids and limits the range of dilational screening, reducing the range of self-limitation (although the scaling $\sim \kappa_{\parallel}^{-1}$ is maintained), as shown by the red points in Fig.~\ref{fig:max size}e.  Note that the finite accessible range of a dilational mechanism is itself structure dependent.  For example, a Kagome structure of rigid triangular units exhibits a larger dilational range (4:1) than square-twist (2:1).  Hence, a simple redesign of subunits to flared triangular prismoidal shapes (see SI) exhibits the same frustrated metamembrane behavior (Fig.~\ref{fig:max size}f).  Compared to square-twist assemblies at the same $\kappa_\parallel/\kappa_\perp$ (Fig.~\ref{fig:max size}e), Kagome assemblies exhibit larger self-limiting lengths.  More significantly, the enhanced range of allowed twist suppresses the excluded volume collapse at the center of the trumpets, permitting the softer (mechanism-screened) stresses to accumulate for a larger range of $\kappa_\parallel/\kappa_\perp$ (see SI).
        
        In summary, the introduction of bulk ``floppy'' metamaterial modes into positionally ordered 2D membranes, qualitatively alters their response to geometric frustration resulting from locally preferred non-Euclidean shapes.  In particular, dilational mechanisms alter the power-law nature of accumulating internal stresses, and significantly extend the total amount and size range over which frustrated membranes can realize hyperbolic curvature, crucially in the continuum limit of vanishing subunit size.  In the context of the self-assembly of frustrated structures, these results demonstrate the significant potential to engineer large self-limiting structures through the careful control of not only the ``misfit'' geometry of the subunits~\cite{Lenz:2017, Spivack:2022}, but also the geometry of their binding, which intentionally introduces soft modes that dramatically enhance the range of intra-assembly stress gradients.  Notably, recent studies of assemblies of {\it de novo} engineered protein materials demonstrate the ability to engineer square-twist, auxetic modes into self-assembled nanomaterials~\cite{Alberstein:2018}.  Similarly, the triangular DNA origami subunit designs used in recent works demonstrating the curvature-programmable assembly of spheres~\cite{Sigl:2021} and cylinders~\cite{Hayakawa:2022} with controllable curvatures are suited for potentially realizing Kagome metamembrane assemblies of the type shown in (Fig.~\ref{fig:max size}).  While self-assembly of frustrated subunits has been demonstrated in finite-temperature models, frustrated metamembranes raise new questions about the role of high conformational entropy in ``floppy'' assemblies relative to standard ``non-floppy'' ones, which may perhaps enhance the thermodynamic stability of self-limiting states \cite{Mao:2013}.  Beyond this, a more basic question to resolve is the nature of the collective mechanics of frustrated metamembranes.  While ``non-floppy'' membranes retain their bulk stiffness even in the absence of frustration, the bulk stiffness of a floppy metamembrane vanishes in the absence of frustration and develops a stable state due to the internal stresses generated at finite frustration.  Hence, it remains to be understood what is the effective modulus of a ``floppy'' frustrated metamembrane and what emergent properties vary with frustration itself.

        The authors are grateful to D. Hall for stimulating discussions on this model.  This work was supported by US National Science Foundation through awards NSF DMR-2028885, DMR-239818, the Brandeis Center for Bioinspired Soft Materials, an NSF MRSEC, DMR-2011846 (M.W. and G.M.G.) and NSF DMR-2217543 (S.R. and C.D.S.). Computational studies performed using the UMass Cluster at the Massachusetts Green High Performance Computing Center.
    
\bibliography{references}
\end{document}